\documentclass[aps,twocolumn]{revtex4}
\usepackage{graphics,graphicx}
\usepackage{color}
\usepackage{wrapfig}
\usepackage{enumerate}
\usepackage{amssymb,amsmath}
\usepackage{bm}
\usepackage{mhchem}
\usepackage[normalem]{ulem} 

\begin{document}
\title{Origin of modulated phases and magnetic hysteresis in TmB$_4$}
\author{Keola~Wierschem$^1$, Sai Swaroop Sunku$^1$,  Tai Kong$^2$, Toshimitsu Ito$^3$, Paul C. Canfield$^2$, Christos Panagopoulos$^1$, and Pinaki~Sengupta$^1$}
\affiliation{$^1$Div. of Physics and Applied Physics, School of Physical and Mathematical Sciences, Nanyang Technological University, Singapore 637371}
\affiliation{$^2$Ames Laboratory, U.S. DOE and Department of Physics and Astronomy, Iowa State University, Ames, Iowa 50011, USA}
\affiliation{$^3$National Institute of Advanced Industrial Science and Technology (AIST), Tsukuba, Ibaraki 305-8562, Japan}
\begin{abstract}
We investigate the low temperature magnetic phases in \ce{TmB4}, a metallic quantum magnet on 
the geometrically frustrated Shastry-Sutherland lattice, using co-ordinated experimental and theoretical 
studies. Our results provide an explanation for the appearance of the intriguing fractional plateau in \ce{TmB4}
and accompanying magnetic hysteresis. Together with observation of the bump in the half-plateau, our 
results support the picture that the magnetization plateau structure in \ce{TmB_4} is strongly influenced 
by the zero-field modulated phases. We present a phenomenological model to explain the appearance of the modulated phases and  a microscopic Hamiltonian that captures the complete magnetic behavior of \ce{TmB_4}.

\end{abstract}
\maketitle

\section{introduction}

Frustrated quantum magnets have emerged as a very fertile proving ground for the discovery of new states of matter. The interplay between competing interactions which cannot be optimized simultaneously, external magnetic fields and (in many instances) enhanced quantum fluctuations due to low dimensionality result in a rich variety of ground state phases with unique functionalities that are not found in their non-frustrated counterparts. Notable examples include spin liquid phases on triangular and kagome lattices, spin ice on pyrochlore lattices and the sequence of Hall-like magnetization plateaus on the two-dimensional (2D) Shastry-Sutherland lattice (SSL) \cite{Balents2010,Mila2011}. Indeed, magnetization plateaus  observed in \ce{SrCu2(BO3)2} have attracted great interest  due to their similarity to quantum Hall physics.~\citep{Kageyama1999,Kodama2002,Sebastian2008,Takigawa2013}

More recently, a new family of rare-earth tetraborides, \ce{RB_4} (R=Tm, Er, Ho, Dy, Tb) has been identified that belong to the Shastry-Sutherland family of quantum magnets. Although they share the same magnetic lattice, the phase diagram of the \ce{RB_4} magnets show a distinct behavior from \ce{SrCu_2(BO_3)_2}. In the insulating \ce{SrCu_2(BO_3)_2}, the exchange interaction is of the Heisenberg-type but the \ce{RB_4} magnets are metallic and the interaction between the moments is of the Ruderman-Kittel-Kasuya-Yosida (RKKY)-type. The RKKY-type interaction, mediated by the conduction electrons, is longer range.
This results in the \ce{RB_4} compounds having a rich phase diagram with multiple long-range modulated phases arising from the competing interactions on a longer length scale.

Among the multiple members of the family, the magnetic properties of \ce{TmB_4} have been investigated most extensively \cite{Gabani2008, Iga2007, Siemensmeyer2008, Michimura2009}. \ce{TmB_4} crystallizes in a tetragonal lattice with the space group {\em P4/mbm} (127). The magnetic moment carrying rare earth ions are arranged in a SSL configuration with weak (magnetic) coupling between the 2D layers. At zero magnetic field, two amplitude-modulated phases have been reported and the Neel state is stable at low temperatures \cite{Michimura2009}. Low temperature field dependent magnetization
data is dominated by a stable plateau at $M/M_{sat}=1/2$ \cite{Iga2007, Gabani2008}. Multiple fractional plateaus such as $M/M_{sat} \sim 1/7, 1/8, 1/9, 1/11\ldots$ accompanied by hysteresis have been reported whose observation has varied between experimental runs \cite{Siemensmeyer2008}.

The sequence of field-induced magnetization plateaus bears striking resemblance to similar behavior in inter-metallic magnets such as \ce{TbNi_2Ge_2}~\citep{Budko1999}, \ce{DySbAg_2}~\citep{Myers1999a,Myers1999b}, and  \ce{HoNi_2B_2C}~\citep{Canfield1997}.
The magnetization plateaus in these metallic quantum magnets result from multiple field-driven metamagnetic transitions arising from competing long range interactions between the localized moments on the rare-earth ions mediated by conduction electrons. However, they differ in one crucial aspect - in \ce{RB4} magnets -- as in other Shastry-Sutherland compounds -- the plateau sequences are primarily determined by the frustrated geometry of the magnetic lattice, rather than purely by RKKY interactions. This is evident from the observation of magnetization plateaus in the insulating \ce{SrCu_2(BO_3)_2} where the only interactions between the localized moments is through superexchange. In case of 
the \ce{RB_4} compounds, both the frustrated geometry and longer range RKKY interactions play a role in the appearance of magnetization plateaus.

The principal features of the ground state magnetic behavior of \ce{TmB_4} were explained by a generalization of the canonical Shastry-Sutherland model. 
A strong crystal electric field lifts the degeneracy of the local spin states. Experimental evidence suggests that the lowest energy state for individual \ce{Tm^{3+}} ions is the 2-fold degenerate non-Kramers doublet $J_i^z = \pm 6$ 
with a large energy gap ($\sim 100K$) to the next energy states.  Accordingly, a
single-ion anisotropy term of the form $-D\sum_i(J_i^z)^2$ is added to the magnetic Hamiltonian. Phenomenologically, the magnitude of the single-ion anisotropy is
estimated to be $D\sim 10K$. Indeed, given the large gap, an effective low energy model comprised of the lowest doublet (with large Ising-like exchange anisotropy) is sufficient to capture the low temperature magnetic behavior.\citep{Iga2007,Gabani2008,Suzuki2009}
It was found that the canonical Shastry-Sutherland interactions need to be supplemented by longer range interactions to capture the observed field dependence. In particular, a fourth-neighbor ferromagnetic interaction is necessary to stabilize a $m/m_s=1/2$ plateau. Two competing explanations have been proposed for the appearance of the fractional plateaus. Siemensmeyer \emph{et. al.} speculated that the fractional plateaus might be magnetic analogues of the plateaus observed in fractional quantum Hall effect, similar to the model proposed for \ce{SrCu2(BO3)2} \cite{Siemensmeyer2008}. Based on a careful neutron scattering study, Michimura \emph{et. al.} proposed a competing explanation in which the fractional plateaus arise directly as an effect of the modulated phases \cite{Michimura2009}. The origin of the modulated phases themselves has also remained unexplained.

In this work, we report a coordinated experimental and theoretical investigation of \ce{TmB_4} aimed at understanding the fractional plateau behavior and the modulated phases. Careful magnetization measurements around the fractional plateau region revealed that the fractional plateau can appear over a continuous value of magnetization around $m/m_s \sim 1/8$, unlike the exact fractions reported previously. We also find that the half-plateau is not completely `flat' but instead has a jump of roughly $m/m_s \approx 1/80$, corresponding to the existence of the second modulated phase. Both of these observations support the model of Michimura \emph{et. al.} for the origin of the fractional plateaus. We also argue that this model leads to a natural explanation of the hysteresis observed at the fractional plateau. We then develop a phenomenological axial next-nearest-neighbor Ising (ANNNI) model to explain the emergence of the modulated phases at zero magnetic field. Finally, we propose a set of parameters for an effective low energy microscopic model that captures the complete magnetic behavior of \ce{TmB_4}.

\section{Experimental details}
Our experiments were performed on TmB$_4$ single crystals synthesized by the solution growth method using an Al solution. Bulk starting elements with a ratio of Tm:B:Al = 0.125 : 0.75 : 50 were put into an alumina crucible, which was heated up to 1475$^{\circ}$C and slowly cooled down to 750$^{\circ}$C in a continuous flow of high-purity argon atmosphere and then quenched to room temperature via furnace cooling. The growth was then taken out from the furnace at room-temperature and re-sealed into a silica ampule. Single crystals of \ce{TmB_4} were separated from the remaining liquid in a centrifuge after heating the ampule back up to 750$^{\circ}$C.

X-ray diffraction in the Laue geometry was used to orient the crystals with an error of less than $\pm 5^{\circ}$. Magnetization measurements were performed in a Quantum Design MPMS XL SQUID magnetometer with the magnetic field along c-axis. 

\section{Experimental results}

\begin{figure}
\centering
\includegraphics[clip,trim=0cm 0cm 0cm 0cm, width=\linewidth]{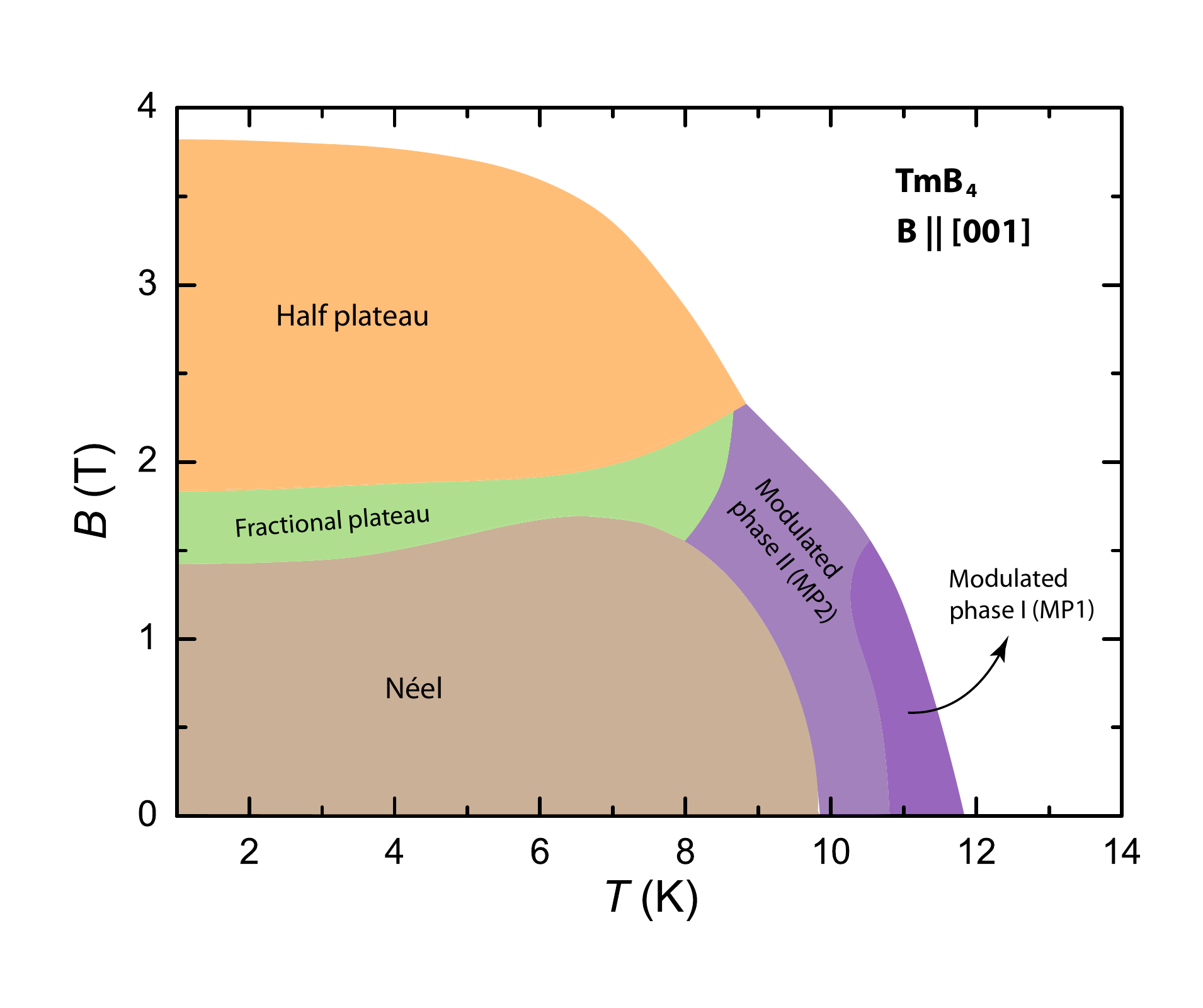}
\caption{(color online) Magnetic phase diagram of \ce{TmB_4} in the $B-T$ parameter space. For small applied fields, the  N{\' e}el state is separated from the high temperature paramagnetic phase by two intermediate phases with modulated magnetic order. At low temperatures, the Ne{\' e}l  order persists to $B\sim 1.4$ T. For higher field strengths,two magnetization plateaus 
are stabilized -- a fractional plateau at  $m/m_s \sim 1/8$ for $1.4 T < B \leq 1.8T$ and a half plateau at $m/m_s =1/2$ for $1.8 T \leq B \leq 3.5 T$.
}

\label{expt_phase}
\end{figure}

The magnetic phase diagram of \ce{TmB4} is well-known from previous studies.~\citep{Siemensmeyer2008,Michimura2009} At small applied fields ($B\lesssim 1.4$T), there is a 3-step thermal transition to a long range ordered Ne{\' e}l state via two 
amplitude modulated AFM phases. Magnetic order sets in at $T_{N1}= 11.8$ K while at $T_{N2}=9.8$ K 
there is a a transition to the Ne{\' e}l state. The two different amplitude modulated
phases are separated by a transition at $T^*=10.9$ K that is visible in resistivity measurements. We identify the phases as follows: Above $T_{N1}$, no long range magnetic order is present. At 
lower temperatures, two amplitude modulated antiferromagnetic (AFM) phases appear. Between 
$T_{N1}$ and $T^*$ Mod. phase I (MP1) in Fig. \ref{expt_phase} -- the amplitude modulation is
indexed by two vectors: $\mathbf{k}_1 = [1\pm k' \pm k'', \pm k'', 0], 
\mathbf{k}_1' = [1\pm k' \pm 3k'', \pm k'', 0]$ $(k' \simeq 0.13, k'' \simeq 0.012)$.~\cite{Michimura2009}
This corresponds to modulations of periodicity of roughly 8 unit cells and 80 unit cells respectively.
Between $T^*$ and $T_{N2}$  -- Mod. phase II (MP2) in Fig. \ref{expt_phase} -- the amplitude
modulation can be indexed by a single vector $\mathbf{k}_2 = [1\pm k', 0, 0]$ $(k' \simeq 0.13)$, 
corresponding to a modulation of roughly 8 unit cells. The transitions at $T_{N1}$ and $T_{N2}$ are also 
visible as anomalies in $dM/dT$, as reported previously.~\citep{Siemensmeyer2008}
The origin of modulated phases Mod. phase I and Mod. phase II has remained unexplained until now.
As we demonstrate below, these zero field modulated phases are crucial for understanding the origin of the
fractional plateau and the accompanying magnetic hysteresis.

\begin{figure}
\centering
\includegraphics[clip,trim=0cm 0cm -0.5cm 0cm, width=\linewidth]
{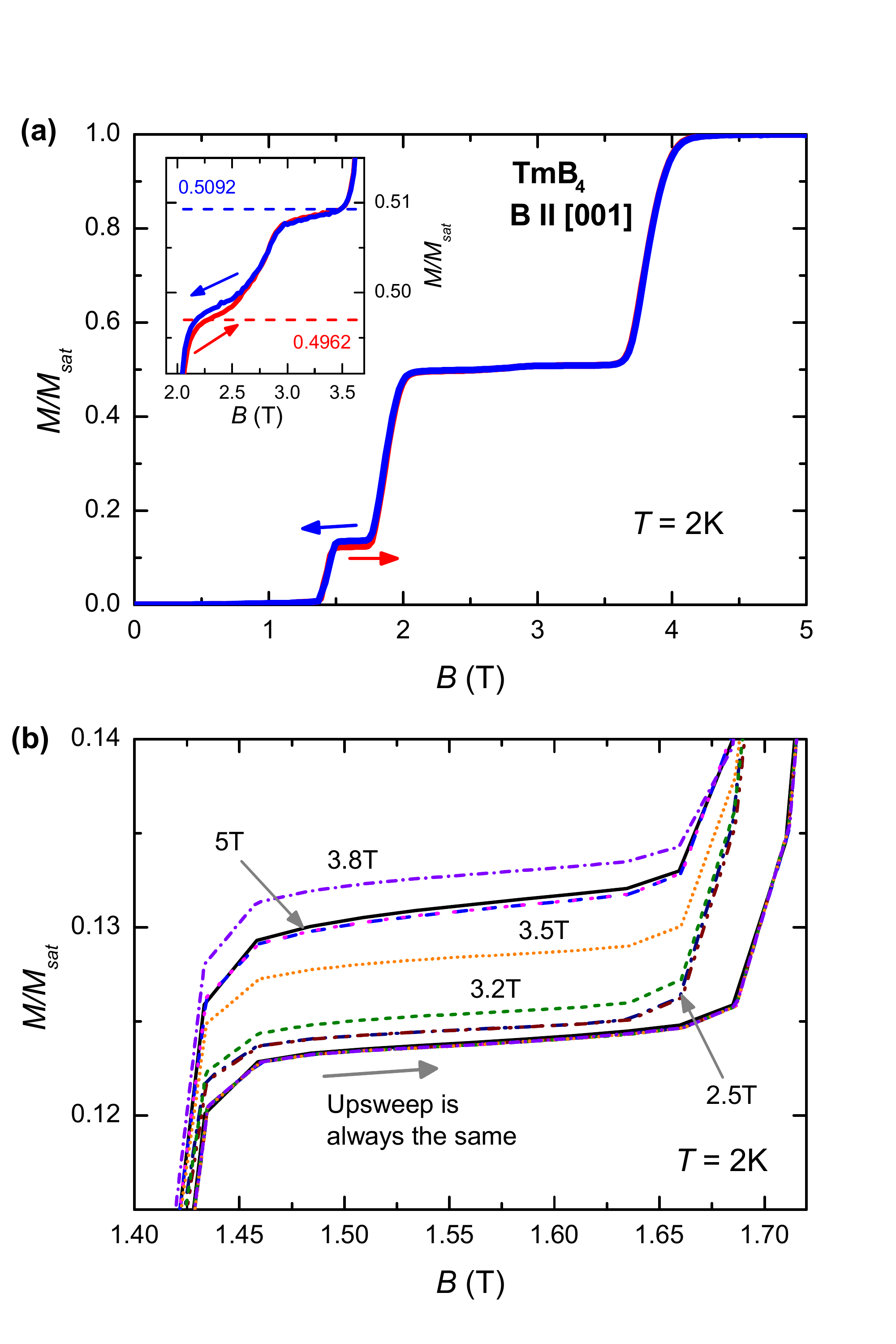}
\caption{(color online) (a) Magnetization of TmB$_4$ along the c-axis normalized to the saturation magnetization at T = 2K. The inset shows a zoom of the bump at the half plateau. (b) magnetization at the fractional plateau for various stopping fields (obtained by using the protocol described in the main text).}
\label{expt_magn}
\end{figure}

When a magnetic field is applied, the fractional plateau phase and the half plateau are stabilized below $\sim$ 9K. Figure \ref{expt_magn}(a) shows the magnetization at 2K. The fractional plateau appears between 1.4T and 1.8T and the half-plateau appears between 2T and 3.5T. As reported previously \cite{Gabani2008,Siemensmeyer2008}, a significant hysteresis is present at the fractional plateau. Upon a closer examination of the half plateau (inset at top left), we find that the latter is not completely `flat', but shows a bump at roughly half its width. We note that this feature is present in previously published data \cite{Siemensmeyer2008,Iga2007,Gabani2008} but has not been discussed so far. This feature gives a finite `height' to the half-plateau of $\simeq 0.013$ $M_{sat} \simeq 1/80$ $M_{sat}$.

It has also been reported that the magnitude of the magnetization at the fractional plateau can vary which has been interpreted as the presence of multiple distinct plateaus. To investigate this phenomena, we performed magnetization measurements according to the following protocol:
\begin{itemize}
\item Cool down to measurement temperature in zero field from above $T_{N1}$
\item Sweep the magnetic field up to 5T to reach the saturation phase and sweep down to 0T
\item Measure while sweeping the magnetic field up to a certain stopping field $B_{stop}$, and measure while sweeping down to 0T from $B_{stop}$.
\end{itemize}

Figure \ref{expt_magn}(b) shows the magnetization at the fractional plateau for various values of $B_{stop}$. The value of the magnetization at the plateau is always the same during the upsweep. The value of the magnetization during the downsweep is always higher than the value during the upsweep but strongly depends on $B_{stop}$. These results indicate that the value of the magnetization at the fractional plateau strongly depends on the exact field history before the measurement. The fractional plateau can appear at almost any value of magnetization during the sweep, as opposed to the fractions 1/7, 1/8, 1/9, 1/11~\ldots reported previously, reflecting the absence of any form of quantization at the fractional plateau. Therefore, our results support the model proposed by Michimura \emph{et. al.} for the origin of the fractional plateau.

\section{modulated structures and plateau hysteresis}

\begin{figure}
\centering
\includegraphics[clip,trim=0cm 8cm 0cm 8cm,width=\linewidth]{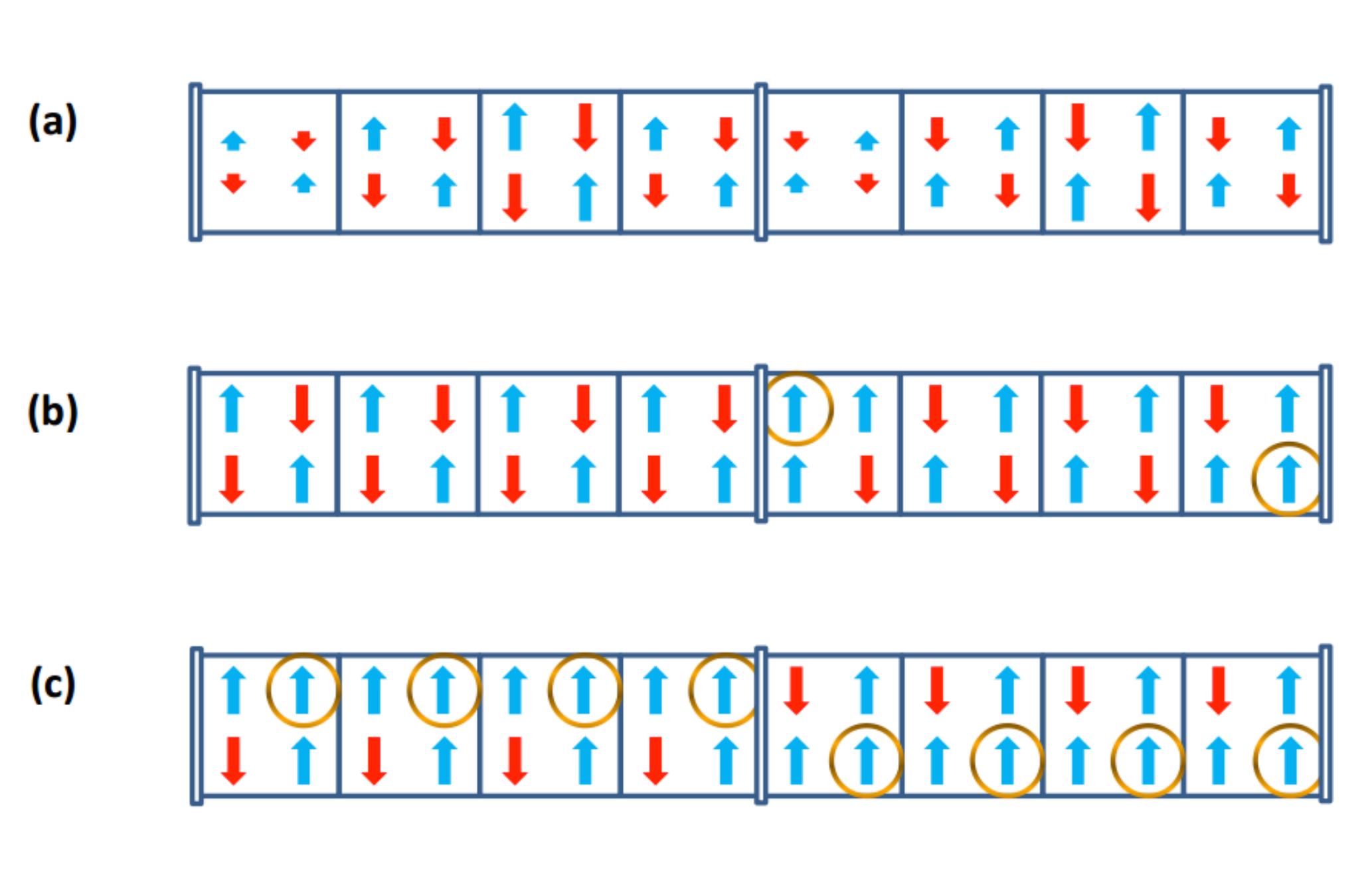}
\caption{(color online) Modulated spin configurations: {\bf (a)}  zero field amplitude modulated intermediate phase as determined by neutron scattering {\bf (b)} possible commensurate magnetic plateau derived from (a) in magnetic field at low temperature {\bf (c)} possible plateau at 1/2 saturation magnetization. We use the same convention as Ref.[\onlinecite{Michimura2009}]for depicting the spins in a unit cell.}
\label{structure}
\end{figure}

We now describe how this model can lead to a natural explanation of the hysteresis at the fractional plateau. Figure~\ref{structure}(b) shows a schematic of the magnetic structure in the modulated phase MP2 as determined by neutron scattering \cite{Michimura2009}. It consists of 4 unit cells of AFM order followed by 4 additional unit cells of AFM order, but with sublattice magnetization reversed. Michimura \textit{et. al.} \cite{Michimura2009} have suggested how this structure can naturally lead to a fractional plateau. In a magnetic field, the spins at the nodes of the amplitude modulated structure are free to align with the field, leading to a paramagnetic contribution. At low temperatures, these ``free'' spins will freeze into an amplitude modulated structure such as Fig.~\ref{structure}(c), leading to a magnetic plateau with fractional magnetization $M/M_{sat} \approx 1/8$, as the periodicity of the modulated structure is 8 unit cells.

The modulated structure shown in Fig.~\ref{structure}(b) is likely only one of several possibilities. 
This is because the the modulation vector in MP2 is incommensurate. While this incommensurability 
does not cause problems in the phase MP2 as it has partially paramagnetic components, the low 
temperature plateau phases will prefer to ``lock on'' to a commensurate magnetic structure. As 
such, any nearby wave vector will do. This can lead to an explanation of the hysteresis of the 
fractional plateaus in TmB$_4$ as follows. During the up-sweep, the fractional plateau regime 
is reached at the lower critical field $H_{c1} (\approx 1.4$ T) at which point a commensurate wave vector 
$k_{com}^1$ is chosen. This wave vector will likely be slightly larger than the incommensurate 
wave vector. If we posit that a change between commensurate wave vectors is akin to a first 
order transition, the system will stay in this plateau until it transitions to the half plateau at higher 
fields. On the way down, the fractional plateau regime is reached at the upper critical field 
$H_{c2} (\approx 1.8$ T) and we can safely assume that a higher magnetization will be favored here as 
$H_{c1}<H_{c2}$. Thus by choosing a commensurate wave vector slightly smaller than the incommensurate one we end up with a fractional magnetic plateau state above the one on the way up, completing our description of magnetic hysteresis in TmB$_4$. 

Next, we consider the effect of the other modulated phase MP1 with an additional modulation of roughly 80 unit cells along the $b$-axis. This additional modulation is expected to influence the low temperature phase at higher fields, i.e. in the 1/2 plateau regime. Thus, we can expect that the 1/2 plateau may have a commensurate version of the above structure, with an extra line of polarized spins every 80 unit cells along the $b$ axis. This may account for an experimentally observed ``bump'' in the 1/2 plateau whose magnitude is roughly 1/80 $M/M_{sat}$.

To summarize, on the way up, modulation begins in the 1/8 plateau, with a structure similar to the intermediate temperature phase but with domain-wall-like lines polarized. At the upper regions of the 1/2 plateau, this amplitude modulation acquires an additional modulation of period 80 along the $b$ axis. Polarization of spins along the domain-wall-like lines leads to a ``bump'' in the 1/2 plateau. On the way down, the additional period 80 modulation disappears at the half-plateau. However, a larger wave vector is selected at the fractional plateau region, leading to hysteresis at the fractional plateau. We have demonstrated that the fractional plateau behavior in \ce{TmB_4} can be understood directly from the existence of the modulated phases.

\section{annni model for t\MakeLowercase{m}b$_4$}

In order to understand the modulated structures in \ce{TmB_4}, we develop a phenomenological ANNNI model. The ANNNI model has previously been well-studied and is considered a prototype for systems with modulated phases \cite{Zhang2010}. By applying a sublattice rotation within the 
$ab$ planes, the AFM state can be turned into a uniform ferromagnet (FM), and interactions between 
moments on different sublattices will switch sign (i.e. FM to AFM and AFM to FM). Thus, we can 
effectively describe TmB$_4$ as an Ising ferromagnet (a uniform external 
field becomes a staggered field in this description). To explain the origin of modulated AFM ordering, 
let us consider the effect of next-nearest-neighbor interactions along the $a$-direction. Such 
a system can be described by the Hamiltonian
\begin{equation}
{\cal H}=\sum_{\alpha}\sum_{\langle ij\rangle_\alpha}{\cal J}_{1,\alpha}J^z_iJ^z_j+\sum_{\langle\langle ij\rangle\rangle_\beta}{\cal J}_{2,\beta}J^z_iJ^z_j,
\label{eq:H}
\end{equation}
with $\alpha\in\{a,b,c\}$ and $\beta\in\{a\}$. Here, the diagonal bonds of the SSL
have been incorporated into $J_{1,a}$ and $J_{1,b}$, as have other bonds not explicitly stated---we are only interested in the minimal model for generating modulated structures. $\langle ij\rangle_\alpha$ and $\langle\langle ij\rangle\rangle_\alpha$ describe nearest and next-nearest neighbor pairs along the $\alpha$-axis. As mentioned previously, we take the ${\cal J}_{1,\alpha}$ to be FM, and therefore ${\cal H}$ reduces to the well-known axial next nearest neighbor Ising (ANNNI) model in the case where all ${\cal J}_{1,\alpha}$ are equal. Even in the case where they are different, we can still expect essentially the same behavior as the phases of the ANNNI model are driven by competition between ${\cal J}_{1,\beta}$ and ${\cal J}_{2,\beta}$.

By constructing a mean field theory of ${\cal H}$ based on the response to a fictitious magnetic field, it can be shown that the optimal spin structure will maximize the Fourier transform of the spin interactions.~\cite{Majlis2007,White2007} In the present case this leads to
\begin{equation}
{\cal J}(k_x)=-{\cal J}_{1,a}\cos(k_x)-{\cal J}_{2,a}\cos(2k_x),
\label{eq:H1}
\end{equation}
where we ignore the trivial $k_y$ and $k_z$ dependence which is optimized by FM order along the $y$- and $z$-axes. To find the optimal $k_x$ value we take the derivative,
\begin{equation}
{\cal J}'(k_x)={\cal J}_{1,a}\sin(k_x)+2{\cal J}_{2,a}\sin(2k_x),
\label{eq:H'}
\end{equation}
and find solutions of $J'(Q)=0$. Aside from the trivial solutions of $Q=0$ and $Q=\pi$ we also find $\cos(Q)=-{\cal J}_{1,a}/4{\cal J}_{2,a}$. Now we evaluate the second derivative,
\begin{equation}
{\cal J}''(k_x)={\cal J}_{1,a}\cos(k_x)+4{\cal J}_{2,a}\cos(2k_x),
\label{eq:H''}
\end{equation}
at these points to find $J''(0)={\cal J}_{1,a}+4{\cal J}_{2,a}$ which is a maxima for ${\cal J}_{1,a}+4{\cal J}_{2,a}<0$ or $-{\cal J}_{1,a}>4{\cal J}_{2,a}$ whereas $J''(\pi)=-{\cal J}_{1,a}+4{\cal J}_{2,a}$ which is a minima for FM ${\cal J}_{1,a}$ and AFM ${\cal J}_{2,a}$ and finally $J''(Q)=-4{\cal J}_{2,a}\sin^2(Q)$ which is negative definite for AFM ${\cal J}_{2,a}$ and hence a maxima whenever $|{\cal J}_{1,a}|<|4{\cal J}_{2,a}|$ as required for $\cos(Q)$ and $\sin(Q)$ to be well defined.

Putting the above together, we can see that for FM ${\cal J}_{1,a}$ and AFM ${\cal J}_{2,a}$ we  have $Q=0$ for $|{\cal J}_{1,a}|>|4{\cal J}_{2,a}|$ while for $|{\cal J}_{1,a}|<|4{\cal J}_{2,a}|$ we have a modulated structure as indexed by $\cos(Q)=-{\cal J}_{1,a}/4{\cal J}_{2,a}$. Using the roughly 8 unit cell modulated structure found by Michimura {\it et al.} for the zero-field modulated phase in TmB$_4$~\cite{Michimura2009} we find ${\cal J}_{2,a}/|{\cal J}_{1,a}|\approx0.27$, where we use $Q^{-1}=8/\pi$ as our unit of distance is half that of a full unit cell.

Next we compare directly to results for the ANNNI model, which have been recently computed using high performance simulations~\cite{Zhang2010}. In the finite temperature phase diagram, the modulated phases only survive at intermediate temperatures. Fitting the width of the intermediate modulated phase in the ANNNI model to the width of the intermediate phase observed in TmB$_4$, we arrive at the ratio ${\cal J}_{2,a}/|{\cal J}_{1,a}|\approx0.35$, comparable to our result from mean field theory.

\begin{figure}
\centering
\includegraphics[clip,trim=0cm 8cm 0cm 8cm,width=\linewidth]{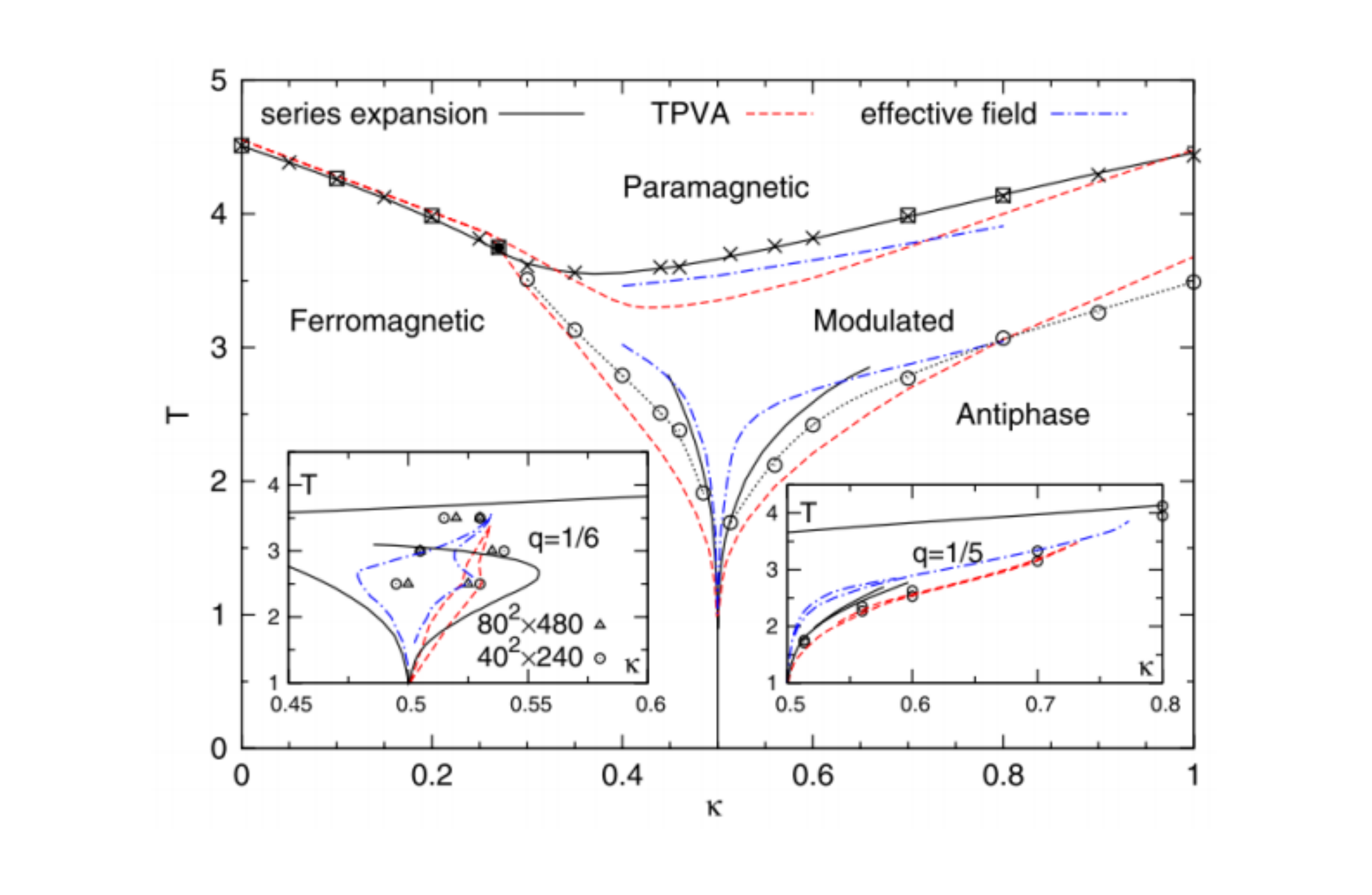}
\caption{(color online) Phase diagram of the ANNNI model, taken from Zhang and Charbonneau~\cite{Zhang2010}. The parameter $\kappa\equiv {\cal J}_{2,a}/|{\cal J}_{1,a}|$.}
\label{phase}
\end{figure}

\section{new t\MakeLowercase{m}b$_4$ parameters}

The dominant features of the field dependence of magnetization in \ce{TmB_4} has previously been explained in terms of a generalized Shastry-Sutherland model~\cite{Suzuki2010}. Most importantly, it was shown that longer range interactions are necessary for the presence of an extended 1/2 plateau and the absence of a 1/3 plateau that is ubiquitous in the Shastry-Sutherland model in the Ising limit.~\cite{Meng2008,Chang2009,Dublenych2012}  However, the parameters introduced by Reference~\cite{Suzuki2010} do not lend themselves to a description of the modulated phases MP1 and MP2, nor do they seem to include the fractional plateau at $m/m_s \sim 1/8$. Huang {\it et al.}~\cite{Huang2013} have considered the parameter set $\{{\cal J}_1,{\cal J}_2,{\cal J}_3,{\cal J}_4\}=\{1,1,0.15,-0.15\}$ for TmB$_4$. Using Glauber dynamics to generate magnetization curves at different field sweep rates, Huang {\it et al.} were able to generate fractional magnetization plateaus near 1/8 of saturation. Similar hysteretic behavior was also seen by Suzuki{\it et al.}~\cite{Suzuki2010} for the parameter set $\{{\cal J}_1,{\cal J}_2,{\cal J}_3,{\cal J}_4\}=\{1,1,0.15,-0.15\}$. However, for the parameter set employed by Huang {\it et al.}, the exact ground state magnetization sequence includes a small but finite width 1/3 plateau~\cite{Wierschem2014}. In addition, the work of Huang {\it et al.}~\cite{Huang2013} fails to capture the fact that the fractional plateau can appear at any value of magnetization.
As such, to develop an accurate microscopic description of the magnetic behavior of \ce{TmB_4}, it is essential to find a consistent set of parameters that lead to an effective ANNNI model. Such a model would exhibit modulated phases at intermediate temperatures, and can also give rise to a fractional plateau at low temperatures under an external field.

The basic requirement for an effective ANNNI model is an AFM next nearest neighbor interaction (${\cal J}_{2,a}$ above). In the works of References~\cite{Suzuki2009,Suzuki2010, Huang2013,Wierschem2013} this term is called ${\cal J}_4$, and the parameters for TmB$_4$ have ${\cal J}_4<0$ i.e. FM. Thus, we see the need to find a new set of parameters if we are to explain the modulated phases of TmB$_4$.

Let us make a list of desirable qualities for any set of parameters describing the interactions between the local moments in TmB$_4$. We will list these in order of importance, and try to keep in mind that the presence of itinerant electrons in TmB$_4$ that mediate the interactions through RKKY-like mechanism may lead to behavior that cannot be explained purely by consideration of local moments alone (for example, the interactions may change as a function of temperature or magnetic field).
\begin{enumerate}
\item{Zero field ground state is the Ising AFM phase with ordering wave vector ${\bf Q}=(\pi,\pi,0)$.}
\item{Width of 1/2 plateau is roughly half of the saturation field.}
\item{No 1/3 or any other plateau except 1/8 and 1/2 plateaus.}
\item{Effective ANNNI model to explain modulated phase at intermediate temperature as well as modulation in 1/8 and 1/2 plateaus.}
\item{1/2 plateau should consist of stripes, as any diagonal or checkerboard arrangement can be excluded according to the neutron scattering analysis of Siemensmeyer {\it et al.}~\cite{Siemensmeyer2008}.}
\item{1/8 plateaus seem to be metastable, and may be due to longer range interactions than can be reasonably considered. Hence, we can always argue they will emerge for any parameter set once longer range interactions are also taken into account.}
\end{enumerate}

\begin{figure}[t]
\centering
\includegraphics[clip,trim=4cm 4cm 4cm 4cm,width=\linewidth]
{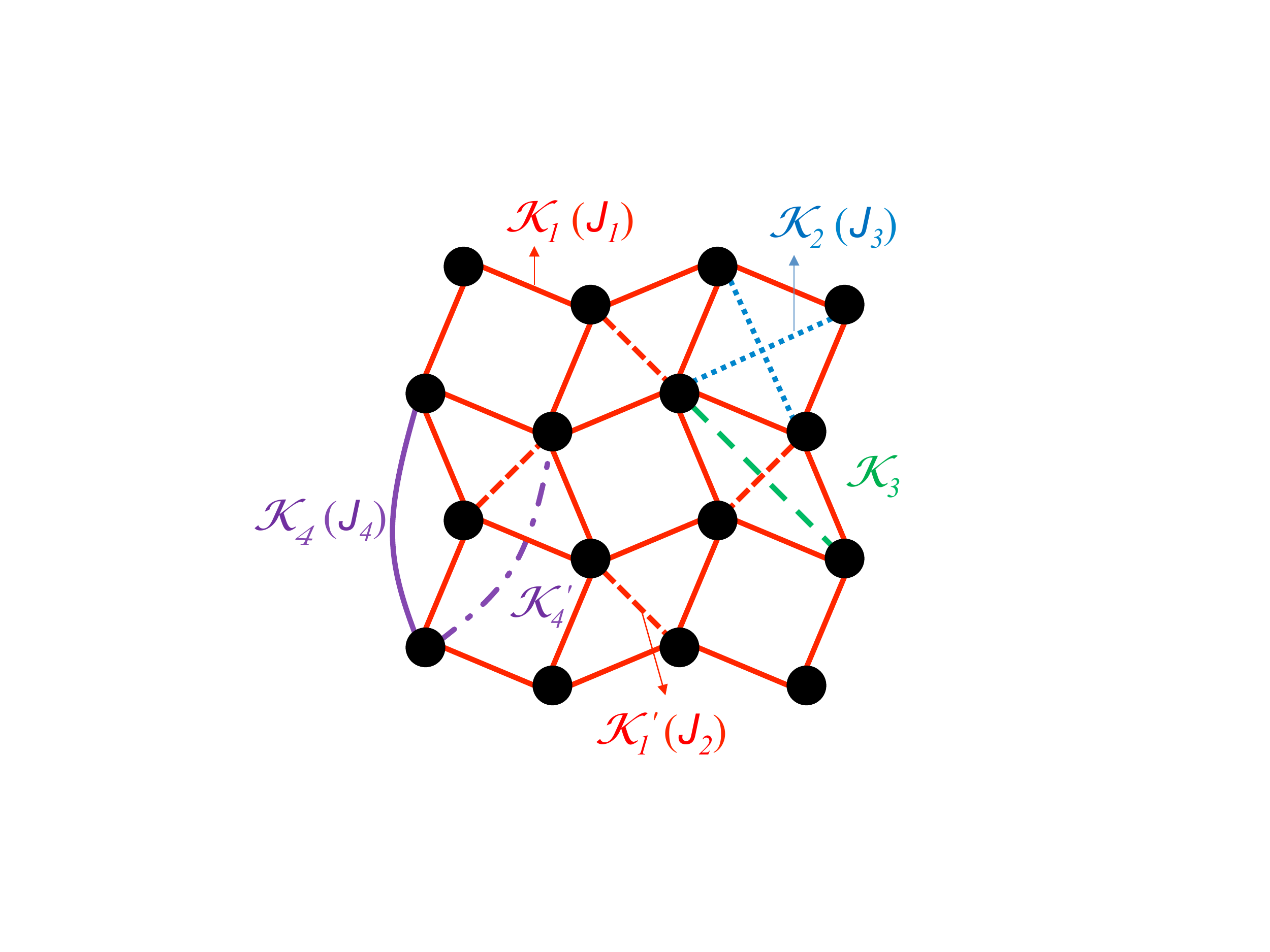}
\caption{(color online) Schematic of the interactions between localized moments in TmB$_4$. Magnetic Tm$^{3+}$ ions are represented by black circles.There are two nearest neighbor interactions with identical bond lengths: ${\cal K}_1$ (solid red lines) and ${\cal K}_1^{'}$ (dashed red lines). Also shown are ${\cal K}_2$ (dotted blue lines) and ${\cal K}_3$ (long dashed green lines), ${\cal K}_4$ (solid purple line) and ${\cal K}_4^{'}$ (dashed purple line). Not all ${\cal K}_2$, ${\cal K}_3$, ${\cal K}_4$, ${\cal K}_4^{'}$ interactions are shown for clarity.}
\label{hierarchy}
\end{figure}

We first tried to achieve the top four qualities by fine-tuning the model used by References~\cite{Suzuki2009,Suzuki2010,Huang2013,Wierschem2013}. In particular, in order to derive an effective ANNNI model, we will have to search for a solution with ${\cal J}_4>0$, in contrast to these works. Using a brute force search through parameter space, we compare the ground state energies of the various known plateaus to find a suitable parameter set that satisfies the first four requirements of the above list. We find that $\{{\cal J}_1,{\cal J}_2,{\cal J}_3,{\cal J}_4\}=\{1,-0.48,-0.93,0.46\}$ works well. However, this has a 1/2 plateau with diagonally arranged stripes, which seems to be precluded by the work of Siemensmeyer {\it et al.}~\cite{Siemensmeyer2008} (point 5 above).

To stabilize a striped 1/2 plateau with a purely horizontal and/or vertical stripe pattern, we need to take into account additional interactions. A natural extension is to consider all possible interactions between the magnetic Tm$^{3+}$ ions in TmB$_4$ up to an interionic cutoff distance equivalent to the ${\cal J}_4$ interaction mentioned above. To do so, we approximate the lattice as an ideal square snub tiling, which is very nearly the case as experimentally determined by x-ray structural determination. In Fig.~\ref{hierarchy} we show a schematic of the interactions, noting that ${\cal J}_1$ and ${\cal J}_2$ as used above form equivalent bond lengths. Thus, in our new parameterization, we refer to them as ${\cal K}_1$ and ${\cal K}_1^{'}$. The next nearest interaction is ${\cal K}_2$ which corresponds to ${\cal J}_3$ above, while ${\cal J}_4$ remains as ${\cal K}_4$. Finally, an additional ${\cal K}_4^{'}$ is possible. Note that for equivalent length bonds, we have used the same subindex, but add a prime to distinguish bonds that may have different character due to the crystal symmetry involved (for example, ${\cal K}_1$ bonds proceed between one octahedral and one dimer Boron, while ${\cal K}_1^{'}$ bonds proceed through two dimer Borons~\cite{Yin2008}).

Within this extended parameter space, our brute force search finds a possible solution for $\{{\cal K}_1,{\cal K}_1^{'},{\cal K}_2,{\cal K}_3,{\cal K}_4,{\cal K}_4^{'}\}=\{1,-0.48,0.44,0.12,-0.12,-0.32\}$. It is interesting to note that in all sets of parameters which lead to an effective ANNNI model, we find ${\cal K}_1^{'}$ (or ${\cal J}_2$ in the model of References~\cite{Suzuki2009,Suzuki2010,Huang2013,Wierschem2013}) to be FM. Note that an ANNNI model is only possible when ${\cal K}_4>{\cal K}_4^{'}$. For comparison, the various parameter sets proposed for TmB$_4$ are listed in Table~\ref{parameters}.

 Recent transport measurements in our group have shown that the c-axis resistivity in
\ce{Tm B_4} is comparable to the in-plane resistivity.\citep{Sunku2015} This is consistent
with previous measurements on Fermi surface in other members of the \ce{RB_4} family. 
In other words, while the magnetic lattice is layered, the electronic transport does not exhibit
such strong anisotropy. This will result in magnetic coupling between the layers mediated
by the conduction electrons. However, such inter-planar RKKY interaction will be much weaker than the
dominant intra-planar exchange interactions between the \ce{Tm^{3+}} ions. We plan to investigate the
effects of itinerant-electron mediated inter-layer magnetic coupling in the future.  


\begin{table*}
\caption{Comparison of critical couplings from various calculations.}
\begin{tabular}{ c c c c c c c }
\hline \hline 
& ${\cal K}_1$ & ${\cal K}_1^{'}$ & ${\cal K}_2$ & ${\cal K}_3$ & ${\cal K}_4$ & ${\cal K}_4^{'}$ \\
& (${\cal J}_1$) & (${\cal J}_2$) & (${\cal J}_3$) & & (${\cal J}_4$) & \\
\hline 
Suzuki {\it et al.}~\cite{Suzuki2010} & 1 & 1 & 0.1182 & 0 & -0.251 & 0 \\
Huang {\it et al.}~\cite{Huang2013} & 1 & 1 & 0.15 & 0 & -0.15 & 0 \\
Present work (diagonal stripes) & 1 & -0.48 & -0.93 & 0 & 0.46 & 0 \\
Present work (vertical stripes) & 1 & -0.48 & 0.44 & 0.12 & -0.12 & -0.32 \\
\hline \hline
\end{tabular}
\label{parameters}
\end{table*}

\section{Summary}

We have conducted a co-ordinated experimental and theoretical investigation of the magnetic
properties of the geometrically frustrated quantum magnet, \ce{TmB_4}, focusing on the 
unusual fractional magnetization plateau and the accompanying magnetic hysteresis. Our key
experimental result is the absence of exact quantization at the fractional plateau. The precise 
value of the magnetization at the fractional plateau and the magnitude of hysteresis depend 
strongly on the field-history. We also observe a bump in the half-plateau that we attribute to 
the presence of the second modulated phases. Both of these results support the model by 
Michimura \emph{et. al.} for the origin of the fractional plateau. We then argue that this model 
leads to a natural explanation of the hysteresis at the fractional plateau.
On the theoretical front, we have developed an effective ANNNI model to describe the 
modulated zero-field AFM phase observed in neutron scattering experiments \cite{Michimura2009}. 
We show that the occurrence of 
the fractional plateau, its variable magnetization and the hysteresis all follow
naturally from the (incommensurate) modulated phase which is explained adequately by an 
effective ANNNI model. Along with a microscopic mechanism for the magnetic hysteresis, our 
results provide specific predictions for the local spin configuration of the
fractional plateau. We hope this will encourage future neutron scattering studies of the fractional 
plateau.  Finally we derive a microscopic Hamiltonian for \ce{TmB_4} that
captures all the observed magnetic behavior, including the magnetization plateaus and 
hysteresis.

\section{Acknowledgment}
It is a pleasure to thank Sriram Shastry and Cristian Batista for useful discussions.
Work in Singapore was supported by grant MOE2011-T2-1-108 from the Ministry of 
Education, Singapore. Work done at Ames Laboratory (PCC and TK) was supported by the U.S. Department of Energy, Office of Basic Energy Science, Division of Materials Sciences and Engineering. The research was performed at the Ames Laboratory. Ames Laboratory is operated for the U.S. Department of Energy by Iowa State University under Contract No. DE-AC02-07CH11358.

\bibliographystyle{apsrev}
\bibliography{modulated-phases-ref}

\end{document}